\documentclass[letterpaper,11pt]{article}   

\usepackage{graphicx}
\usepackage{geometry}

\begin{document}

\title{Actively stabilized wavelength-insensitive carrier elimination from an electro-optically modulated laser beam}


\small{\author{Nathan Cooper,$^{1,*}$ James Bateman,$^1$ Alexander Dunning$^1$ and Tim Freegarde$^{1}$\\
$^1$School of Physics and Astronomy, University of Southampton, \\ Highfield, Southampton, Hampshire SO17 1BJ, UK\\
$^*$Corresponding author: nlc2g09@soton.ac.uk}}

\maketitle

\begin{center}
 \small{Published in J. Opt. Soc. Am. B / Vol. 29, No. 4 / April 2012 / pp. 646-649\\
 Copyright 2012 Optical Society of America}
\end{center}

\begin{abstract}
We demonstrate a simple and robust technique for removal of the carrier wave from a phase-modulated laser beam, using a non-interferometric method 
that is insensitive to the modulation frequency and instead exploits the polarization-dependence of electro-optic modulation. An actively stabilized 
system using feedback via a liquid crystal cell yields long-term carrier suppression in excess of 28~dB at the expense of a 6.5~dB reduction in sideband power.
\end{abstract}

\section{Introduction}

Laser stabilization techniques and studies in laser physics often require 
the generation of frequency shifted radiation that is otherwise coherent with light from a master laser source, as do many experiments
in atomic or molecular physics --- particularly those which address hyperfine atomic structure \cite{spectrometer, raman, offsidlock}.  
This is typically achieved with an acousto-optic (AOM) \cite{MOIL} or electro-optic modulator (EOM) \cite{injlock}, 
or via current modulation of a semiconductor laser \cite{currentmod}. It is usually necessary to separate the frequency-shifted radiation from the unshifted carrier wave. 
While trivial with an AOM, carrier extraction is less straightforward when the modulation frequency merits use of an EOM, and several techniques have been developed for this purpose \cite{faradaysplit, machzen, injlock2}.

Established methods generally operate by separating light according to its wavelength, and have two major drawbacks. 
Firstly, interferometric methods require sophisticated and costly apparatus to stabilize a resonant cavity or other optical path with sub-wavelength accuracy \cite{machzen,injlock2} 
and a vibration-free environment is essential. Secondly, wavelength-dependent techniques are inappropriate if rapid changes in modulation frequency are to be accommodated. Indeed, many schemes are only able 
to operate at a single modulation frequency \cite{faradaysplit, machzen}. 
\newpage
With electro-optic modulation, the sidebands may in principle be at least partially separated from the carrier using polarization techniques,
as they do not necessarily share its polarization state, 
but the temperature-dependent birefringence of the modulator necessitates continuous adjustments if carrier extinction is to be maintained. 
Here we present a scheme in which this birefringence is actively compensated by a liquid crystal cell, allowing carrier suppression to be maintained at over 
28~dB, albeit at the expense of a 6.5~dB attenuation of the desired sideband. Our simple and economical approach avoids wavelength-scale optical path stabilization 
and can function continuously irrespective of changes in modulation frequency.

\section{Principles of Operation}

\begin{figure}
  \begin{center}
   \includegraphics[width = 10.0cm]{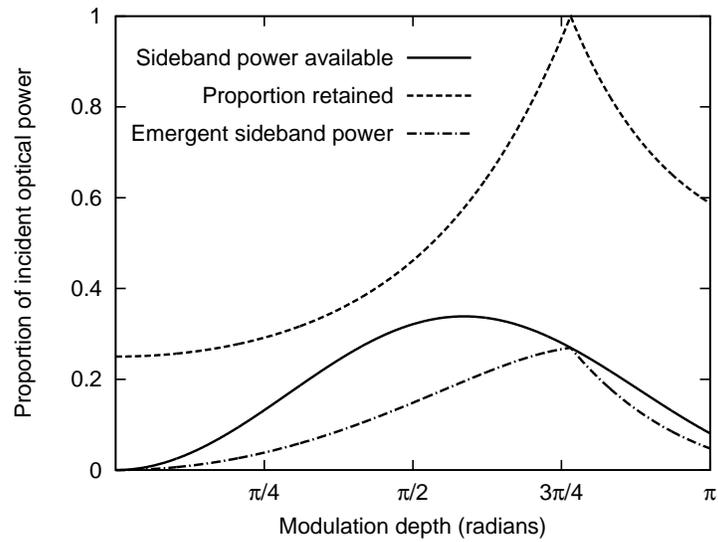}
   \caption{Optimized performance against modulation depth: the three lines plotted correspond to the maximum proportion of the input 
power that could be placed into the 1st order sidebands without
carrier removal, the proportion of this power that can be retained under polarization filtering arrangements and hence the maximum proportion
of the input power that can be placed into carrier-free 1st order sidebands.}
   \label{powerplot}
  \end{center}
\end{figure}

Most electro-optic modulators affect only a single linear polarization component of the incident light field \cite{EOM}. 
We consider an EOM which, driven at a frequency $\Omega$, achieves a modulation depth $m$ \cite{maximumphaseexcursion} 
for the modulated polarization component of incident light with a frequency $\omega$. If the incident beam is linearly polarized \cite{deferproof} with (real)
amplitudes $A \cos{\vartheta}$ and $A \sin{\vartheta}$ in the modulated and unmodulated directions respectively, then the phase-modulated output may be written
\begin{equation}
\mathbf{E} = A \left( \begin{array}{c}   e^{i(\omega t + m \cos \Omega t)} \cos{\vartheta} \\  e^{i \omega t} \sin{\vartheta} \end{array} \right) \, .
\end{equation}
Decomposing the modulated polarization into the resulting frequency components using the Jacobi-Anger identity yields
\begin{equation}
  \mathbf{E} = A \left(\begin{array}{ccc}
    J_{0}(m) \cos{\vartheta} \\
  \sin{\vartheta}   \\
  \end{array} \right) e^{i\omega t}
  + A \left(\begin{array}{ccc}
  \cos{\vartheta} \\
  0 \\
  \end{array} \right) e^{i\omega t} \displaystyle\sum_{n \neq 0}  i^{n} J_{n}(m) e^{i n \Omega t}
,
\end{equation}
where $J_{n}$ is the $n$th order Bessel function of the first kind;
the first term represents the carrier wave and the remaining terms represent the sidebands.   
If the carrier wave is removed by a polarizer aligned so that its transmission axis is orthogonal 
to the polarization plane of the emerging carrier, then the transmitted sidebands will be
\begin{equation}
\mathbf{E}_{sidebands} = \left( \begin{array}{c} -\sin\phi \\ \cos\phi \end{array} \right) A \cos{\vartheta} \sin\phi \sum_{n \ne 0} i^{n} J_{n}(m) e^{i(\omega + n \Omega) t} \, ,
\end{equation}
where $\phi = \arctan \{ \tan{\vartheta} / J_{0}(m)\}$ is the angle between the polarizer's 
transmission axis and the modulated direction. The fraction of the incident power emerging in each of the first-order ($n = \pm 1$) sidebands is therefore
\begin{equation}
P_{S1} =  \cos^{2}{\vartheta} \sin^{2}{\phi} ~J_{1}^{2}(m) = \frac{ \cos^{2}{\vartheta} \tan^{2}\phi}{1+\tan^{2}\phi} J_{1}^{2}(m) = \frac{ \cos^{2}\vartheta J_{1}^{2}(m)}{\{1+[J_{0}(m) \cot{\vartheta} ] ^{2}\} }. 
\end{equation}
This is maximized (for modulation depth $m$) when $\cos^{2}{\vartheta} = 1/(1 + |J_{0}(m)|)$. The system then retains a proportion of the sideband power
that would be available without carrier removal equal to $1/(1 + |J_{0}(m)|)^{2}$, as shown in Fig.~\ref{powerplot}.     

\begin{figure}
  \begin{center}
   \includegraphics[width = 8cm]{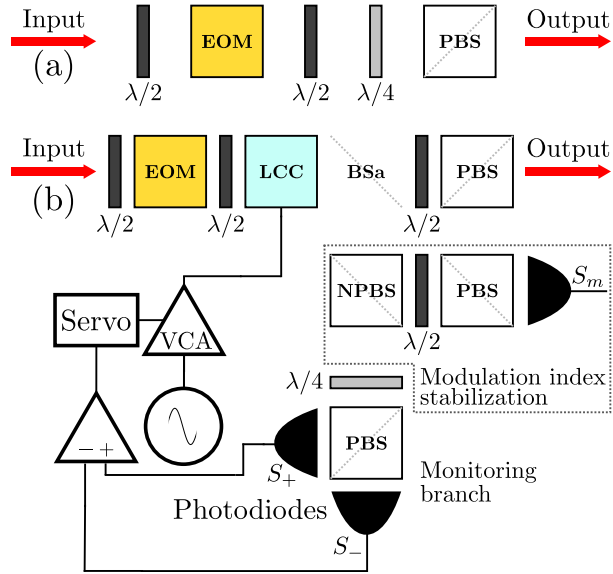}
   \caption{(a) Polarization optics for passive removal of the carrier; the wave plates after the EOM must be adjusted to compensate for any birefringence. 
(b) Equivalent arrangement using a liquid crystal cell for active compensation. The residual birefringence is monitored using a H\"{a}nsch-Couillaud scheme. 
BSa: non-polarizing beam sampler. EOM: electro-optic modulator. LCC: liquid crystal cell. 
(N)PBS: (non-)polarizing beam splitter. VCA: voltage controlled amplifier. In practice the electronic feedback was implemented digitally.}
   \label{diagram}
  \end{center}
\end{figure}

\newpage
The crystals upon which electro-optic modulators are usually based exhibit substantial birefringence, which introduces a phase difference 
between the polarization components in the modulated and unmodulated directions \cite{generalise}, changing the polarization from linear to elliptical
and thereby reducing the attenuation of the carrier wave by the polarizing filter. Additional polarization optics must therefore be 
introduced to counteract this effect. A suitable passive set-up, using a half-wave plate and a quarter-wave plate, is shown in Fig.~\ref{diagram}(a).
However, since the birefringence of the crystal changes with temperature, long-term stability requires either precise temperature stabilization or 
feedback control of the extra polarization components.

To achieve this, we use an E7 liquid crystal cell \cite{whatthehellsE7} as a voltage-controlled wave plate, similar to that discussed in \cite{biref}. 
The arrangement for this is shown in Fig.~\ref{diagram}(b). Such cells can readily be produced via the method detailed 
in \cite{lqdcrystp} and are also available commercially \cite{phaseretarder}. The liquid crystal shows a birefringence which depends upon the amplitude of an AC voltage applied across it; we use a 1~kHz sine wave, and find 
that the phase difference between the two polarizations can be varied (approximately linearly, as apparent in Fig.~\ref{errsig}) by several cycles as the peak-to-peak voltage is increased 
from 1.2~V to 2.6~V. We do not see any modulation of the birefringence at the frequency of the AC signal. 

We generate an error signal by using a quarter-wave plate and polarizing beam splitter to direct opposite circular polarization components onto the pair 
of photodiodes in Fig.~\ref{diagram}(b), as in the popular spectroscopy 
arrangement of H\"ansch and Couillaud \cite{Han-Cou}. This is fed back to the liquid crystal cell through a servo controller that determines the amplitude of the AC modulation \cite{controller}.
With the axes of the quarter-wave plate making angles of $\pi/4$ with the axes 
of the polarizing beam splitter, the two photodiode signals are
\begin{equation}
 S_{\pm} = KA^{2} [ 1/2 \pm \cos{\vartheta} \sin{\vartheta} J_{0}(m) \sin{\delta} ], 
\end{equation}
where $K$ represents the photodiode sensitivity and $\delta$ is the net birefringent phase difference introduced by   
the EOM and the liquid crystal cell. Thus the error signal is given by
\begin{equation}
 S_{E} = S_{+} - S_{-} 
         = 2 K A^{2} \cos{\vartheta} \sin{\vartheta} J_{0}(m) \sin{\delta},
\end{equation}
which has a finite gradient and takes a value of zero at $\delta = 0$, as required for locking of $\delta$ to zero. 
The linearly polarized sidebands are divided equally between the two photodiodes and hence do not contribute to the error signal.

\section{Experimental Results}

\begin{figure}
  \begin{center}
   \includegraphics[width = \textwidth]{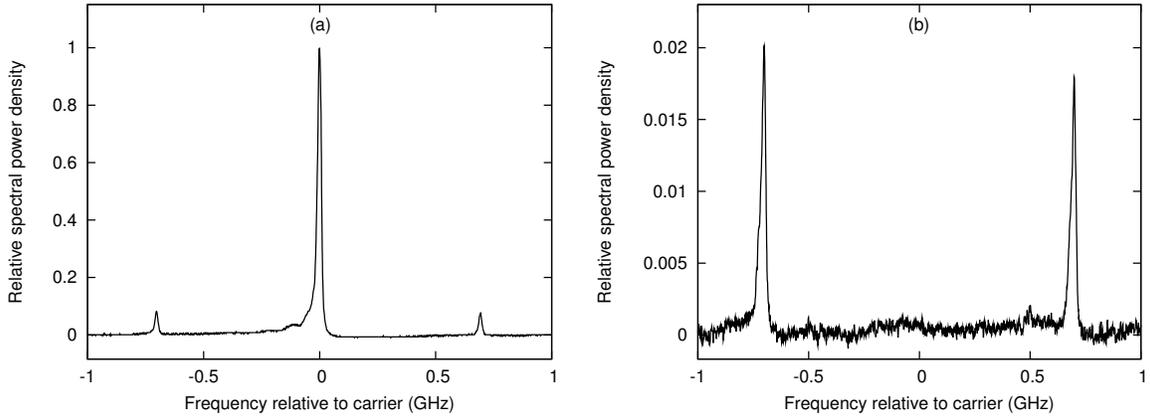}
   \caption{Spectra of light (a) leaving the EOM and (b) emerging from the carrier removal system. The
modulation frequency is 2.7 GHz and our optical spectrum analyzer has a free spectral range of 2 GHz, hence the 
apparent appearance of the first order sidebands at a relative frequency of $\pm$ 700~MHz. The scales are consistent between
the two panels.}
   \label{spectra}
  \end{center}  
\end{figure}

\begin{figure}
  \begin{center}
   \includegraphics[width = 9cm]{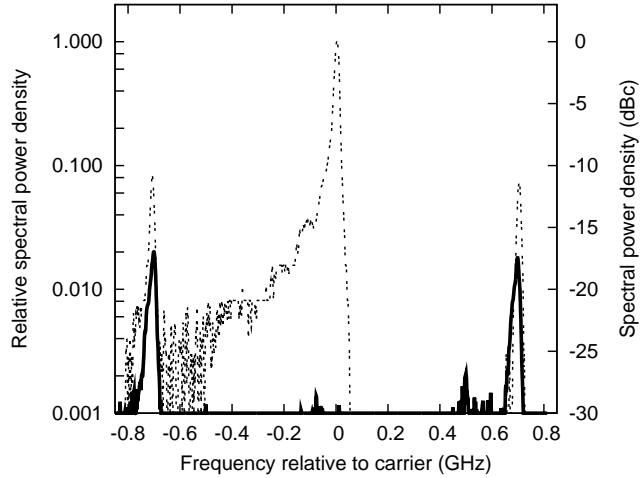}
   \caption{Logarithmic plot of the optical spectra before (dotted lines) and after (solid lines) carrier removal, smoothed with a 20~MHz bandwidth moving average filter. The lower readings of the dotted trace are adversely affected by experimental noise and the finite
resolution of the oscilloscope. The shoulders to the left of the carrier peak are higher order transverse modes resulting from a small misalignment of the optical spectrum analyser.}
   \label{logspec}
  \end{center}  
\end{figure}

Both systems shown in Fig.~\ref{diagram} were constructed and their outputs were observed using an optical spectrum analyser. 
Figs.~\ref{spectra} and \ref{logspec} show spectra of the light produced without (a) and with (b) active carrier removal (we used a modulation depth of $\sim \pi/5$~radians). Similar results could be obtained 
over short periods with the passive arrangement of Fig.~\ref{diagram}(a), giving instantaneous carrier extinction levels in excess of 30 dB 
(consistent with the performance of the polarizing beam splitters), but variations in birefringence within the EOM caused the carrier 
transmission to increase to over 2\% after only five minutes of operation. With the active stabilization of Fig.~\ref{diagram}(b), the carrier 
could be suppressed indefinitely, and we recorded a time-averaged carrier wave suppression over a period of 2.5 hours of 28.8 dB [20]. 
Fig.~\ref{logspec} shows that the sideband powers were attenuated by 6.5 dB, in good agreement with equation (4).

\begin{figure}
  \begin{center}
   \includegraphics[width = 9cm]{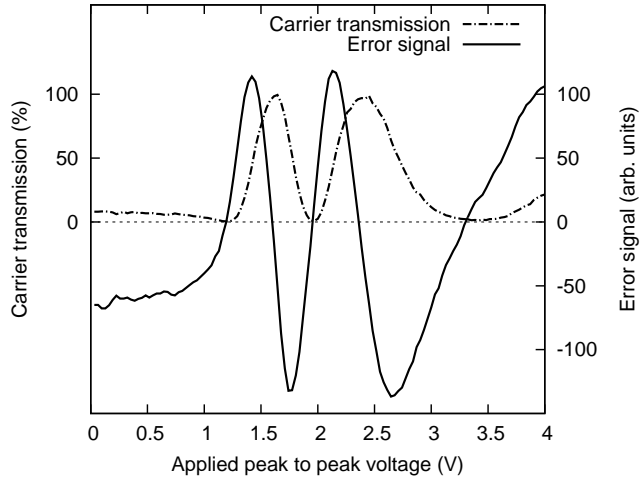}
   \caption{Error signal and carrier transmission (as a percentage of its maximum value) during a sweep of the peak to peak voltage applied to the LCC.}
   \label{errsig}
  \end{center}
\end{figure}

Experimental values of carrier transmission and error signal, $S_{E}$, are plotted against the peak to peak voltage applied to the liquid crystal cell ($V_{pp}$) in Fig.~\ref{errsig}. The behavior of the error signal is 
consistent with that expected theoretically, with $S_{E}$ crossing zero at the minimum of carrier transmission and the variation of the signal with $V_{pp}$ being approximately 
sinusoidal between 1.2~V and 2.6~V. 
There is a small but systematic difference in magnitude between the positive and negative going parts
of the error signal, not predicted by equation (6). We attribute this difference to etalon effects in the (uncoated) liquid crystal cell, with reflections 
at the glass -- liquid crystal interfaces
causing the transmission to vary slightly with changes in the optical path length of the cell.       

\section{Compensation for variations in modulation depth}

Although the lock point of the birefringence compensation is independent of $m$, changes in modulation depth would require realignment of the polarizing beam 
splitter (or in our case the half-wave plate) on the 
output branch  
in order to maintain the condition $\phi = \arctan \{ \tan{\vartheta}/J_{0}(m)\}$. A signal could be generated to control this by splitting the beam on the 
monitoring branch (before the quarter-wave plate) and creating a new branch as shown in Fig.~\ref{diagram}(b). With $\delta$ locked to zero and the modulated
polarization component making an angle of $\pi/4$ with the axes of the polarizing beam splitter,
the signal produced by the photodiode, $S_{m}$, could be normalised to give:
\begin{equation}  
 \hat{S}_{m} = \frac{S_{m}}{(S_{+} + S_{-})} = \frac{K A^{2}[1/2  + \cos{\vartheta} \sin{\vartheta} J_{0}(m) \cos{\delta} ]}{K A^{2}} = \frac{1}{2} +  \cos{\vartheta} \sin{\vartheta} J_{0}(m)  .
\end{equation}
This gives a measure of $J_{0}(m)$ that could be used to adjust the radio frequency power with which the EOM is supplied, in order to maintain a constant modulation
depth. Alternatively it could be used to control a motorized wave plate or other additional polarizing components that would 
allow adjustment of the system to cope with changes in modulation depth.  
 
\section{Conclusions}
We have demonstrated a method for removing the carrier wave from an electro-optically phase-modulated spectrum that is unaffected by 
changes in modulation frequency and both easier to implement and more robust than most existing schemes. It achieves these advantages
at the expense of a substantial loss ($\simeq$75\% at low modulation index) in total power, making it useful when spectral purity is more important than absolute power.
Using a H\"{a}nsch-Couillaud stabilization method, we have demonstrated carrier suppression of over 28 dB, limited mainly by the extinction ratio of the 
polarizing beam splitters, which can be maintained for a number of hours. We have suggested a scheme by which any changes in modulation depth, which 
determines the polarizer alignment, could be monitored and actively compensated.

\section{Acknowledgements}
The authors would like to thank Mark Herrington for supplying the liquid crystal cell.
This work was supported by the UK EPSRC (EP/E039839/1 and EP/E058949/1) and by the CMMC collaboration within the ESF EuroQUAM programme.


\begin{thebibliography}{99}
\bibitem{maximumphaseexcursion} The maximum modulation depth achieveable with most commercial EOMs is of the order of 1~radian.
\bibitem{deferproof} Although we currently assume a linear input polarization, no extra calculation is required to generalise our treatment to any input polarization.
\bibitem{generalise} We now see that any phase relationship between the horizontally and vertically polarized components of the input light can be subsumed into this phase difference, hence our calculations are valid for any input polarization.
\bibitem{controller} This was implemented using an Atmel ATMEGA328P-PU microprocessor (mounted on an Arduino Uno board) and Analog Devices DAC8562 12~bit digital to analog converter.
\bibitem{limitation2} The long-term carrier suppression was slightly reduced by small variations in the modulation depth; with the enhancement of section 4, 
we expect that performance would be limited by the polarizing beam splitters.
\end{thebibliography}
\end{document}